\documentstyle[multicol,aps,epsf]{revtex}

\begin{document}
\title{Fixation in a cyclic Lotka-Volterra model}
\author{L.~Frachebourg$^\ast$ and P.~L.~Krapivsky$\dag$}
\address{$^\ast$Laboratoire de Physique Statistique, Ecole Normale 
Sup\'erieure, F-75231 Paris Cedex 05, France}
\address{$\dag$Center for Polymer Studies and Department of Physics,
Boston University, Boston, MA 02215, USA}
\maketitle
\begin{abstract} 
We study a cyclic Lotka-Volterra model of $N$ interacting species
populating a $d$-dimensional lattice.  In the realm of a Kirkwood
approximation, a critical number of species $N_c(d)$ above which the
system fixates is determined analytically.  We find $N_c=5,14,23$ in
dimensions $d=1,2,3$, in remarkably good agreement with simulation
results in two dimensions.

\medskip\noindent
{PACS numbers:  02.50.Ga, 05.70.Ln, 05.40.+j}
\end{abstract}
\begin{multicols}{2}

A cyclic variant of the Lotka-Volterra model of interacting populations,
originally introduced by Vito Volterra for description of struggle for
existence among species\cite{volterra,montroll}, has then appeared in a
number of apparently unrelated fields ranging from plasma
physics\cite{zakh} to integrable systems\cite{itoh,bog}.  Recently, the
cyclic Lotka-Volterra model (also known as the cyclic voter model) has
attracted a considerable interest as it was realized that introduction
of the spatial structure drastically enriches the dynamics
\cite{bramson,t,f,fg,gr88,lpe}.  Namely, if species live on a
one-dimensional (1D) lattice, a homogeneous initial state evolves into a
coarsening mosaic of interacting species.  This heterogeneous spatial
structure spontaneously develops when the number of species is
sufficiently small, $N<N_c$, where $N_c=5$ in one
dimension\cite{bramson,f,lpe}.  For $N\geq N_c$ fixation occurs,
i.e. the system approaches a frozen state.  Little is known in higher
dimensions, even existence of $N_c$ has not yet been established
theoretically or numerically (in simulations on 2D lattices with $N\leq
10$ species, no sign of fixation has been found and instead a reactive
steady state has been observed \cite{bramson,t,f,fg,gr88,lpe}).  In this
work we investigate the cyclic Lotka-Volterra model in the framework of
a Kirkwood-like approximation.  This approach predicts a {\em finite}
$N_c$ in all spatial dimensions.

In the following, we shall use the language of voter model\cite{lig}.
Consider the cyclic voter model with $N$ possible opinions.  Each site
of a $d$-dimensional cubic lattice is occupied by a voter which has an
opinion labeled by $\alpha$, with $\alpha=1,\ldots ,N$.  Voters can
change their opinions in a cyclic manner, $\alpha\to \alpha-1$ modulo
$N$, according to the opinions of their neighborhood.  Specifically, the
following sequential dynamics is implemented: (i) we choose randomly a
site (of opinion $\alpha$, say) and one of its $2d$ nearest neighbors
(of opinion $\beta$); (ii) if $\beta=\alpha-1$, then the chosen site
changes its opinion from $\alpha$ to $\beta=\alpha-1$; (iii) otherwise,
opinion does not change.  We set the time scale so that in unit time
each site of the lattice is chosen once in average.  When $N=2$ the
cyclic voter model is identical to the classic voter model which is
solvable in arbitrary dimension\cite{pk}; therefore in the following we
assume that $N\geq 3$.

In order to simplify notations, we consider first a 1D chain.  We define
$p_{\alpha_1,\ldots,\alpha_i}(t)$ as the probability that a randomly
chosen segment of $i$ consecutive sites contains opinions
$\alpha_1,\ldots,\alpha_i$.  For instance, the one-point function
$p_\alpha(t)$ is just the density of opinion $\alpha$.  It obeys

\begin{equation}
\label{one}
2{dp_\alpha\over dt}=p_{\alpha,\alpha+1}+p_{\alpha+1,\alpha}
-p_{\alpha,\alpha-1}-p_{\alpha-1\alpha,}.
\end{equation}
We consider random and uncorrelated initial opinion distributions.  This
implies $p_\alpha(0)=1/N$, and generally
$p_{\alpha_1,\ldots,\alpha_i}(0)=1/N^i$.  Symmetry leads to
$p_{\alpha,\alpha+1}=p_{\alpha+1,\alpha}
=p_{\alpha-1,\alpha}=p_{\alpha,\alpha-1}$, so Eq.~(\ref{one}) gives
$dp_\alpha/dt=0$ and hence $p_\alpha(t)=1/N$.  Although the dynamics is
nonconserved, i.e. the densities can change locally, we see that for the
symmetric initial conditions with equal concentrations the densities are
conserved globally.  The two-point functions obey

\begin{eqnarray}
\label{0}
2{dp_{\alpha,\alpha}\over dt}=&-&p_{\alpha-1,\alpha,\alpha}
-p_{\alpha,\alpha,\alpha-1}+p_{\alpha+1,\alpha}\nonumber\\
&+&p_{\alpha,\alpha+1} +2p_{\alpha,\alpha+1,\alpha},
\end{eqnarray}
\begin{eqnarray}
\label{1}
2{dp_{\alpha,\alpha+1}\over dt}=&-&p_{\alpha,\alpha+1}
-p_{\alpha-1,\alpha,\alpha+1}-p_{\alpha,\alpha+1,\alpha}\nonumber\\
&+&p_{\alpha,\alpha+1,\alpha+1}+p_{\alpha,\alpha+2,\alpha+1},
\end{eqnarray} 
which are valid for arbitrary $N\geq 3$, and

\begin{eqnarray}
\label{i}
2{dp_{\alpha,\alpha+i}\over dt}=&-&p_{\alpha-1,\alpha,\alpha+i}
-p_{\alpha,\alpha+i,\alpha+i-1}\nonumber\\
&+&p_{\alpha,\alpha+1,\alpha+i}+p_{\alpha,\alpha+i+1,\alpha+i}.
\end{eqnarray}
Eqs.~(\ref{i}) apply for $N\geq 4$, $2\leq i\leq N-2$.  Of course, the
indices are taken modulo $N$.  Finally, for symmetry reasons
$p_{\alpha,\alpha+N-1}=p_{\alpha,\alpha-1}=p_{\alpha,\alpha+1}$, and
more generally $p_{\alpha,\alpha+N-i}=p_{\alpha,\alpha+i}$.

Above equations are exact and normalization can be verified.  For
instance, $\sum_{1\leq i\leq N}p_{\alpha,\alpha+i}=p_\alpha=1/N$.
Eqs.~(\ref{one})--(\ref{i}) are the first of an infinite hierarchy of
equations which is hardly solvable.  However, a considerable insight can
be gained within the two-sites mean-field approximation (also called
Kirkwood approximation) that expresses $k$-point functions via one- and
two-point functions\cite{balescu}.  For example, the three-point
functions read

\begin{equation}
\label{kirkwood}
p_{\alpha_1,\alpha_2,\alpha_3}=
{p_{\alpha_1,\alpha_2}p_{\alpha_2,\alpha_3}\over p_{\alpha_2}}.
\end{equation}
This kind of factorization approximation originally developed in the
realm of equilibrium statistical mechanics has proven to be remarkably
successful for a number of non-equilibrium processes as well\cite{kirk}.

The ansatz of Eq.~(\ref{kirkwood}) closures the above rate equations;
e.g., Eqs.~(\ref{i}) become
\begin{equation}
\label{dif}
\dot r_i={Nr_1\over 2}\left(r_{i-1}-2r_i+r_{i+1}\right),
\end{equation}
where $r_i=p_{\alpha,\alpha+i}$, so for instance
$r_1=p_{\alpha,\alpha+1}$ is the concentration of reactive pairs.  Note
that the evolution rules which define the model are translationally
invariant in ``opinion space'' and therefore for translationally
invariant initial distributions the two-point correlator
$p_{\alpha,\beta}$ is only a function of $\beta-\alpha$.  Hence $Nr_i$
is the probability that opinions of any two randomly chosen consecutive
sites differ by $i$.  The normalization condition thus reads
$\sum_{0\leq i\leq N-1}r_i=1/N$.  Upon combining with the symmetry
requirement, $r_i=r_{N-i}$, the normalization condition yields
\begin{eqnarray}
\label{eonorm}
r_0+2\sum _{i=1}^{M-1} r_i+r_{M}={1\over N}, \quad N=2M,\\
r_0+2\sum _{i=1}^M r_i={1\over N}, \quad N=2M+1.
\end{eqnarray}

We now turn to the arbitrary dimension $d$.  Making use of the compact
notations $r_i$, we arrive at the generalization to the previous rate
equations (valid within the realm of Kirkwood approximation)
\begin{eqnarray}
\label{main}
\dot r_0 &=& {2d-1\over 2d}Nr_1\left[{2\over
(2d-1)N}-2r_0+2r_1\right], \nonumber\\
\dot r_1&=&{2d-1\over 2d}Nr_1
\left[-{1\over (2d-1)N}+r_0-2r_1+r_2\right],\\
\dot r_i &=& {2d-1\over 2d}Nr_1\left[r_{i-1}-2r_i+r_{i+1}\right],
\quad i=2,\ldots,M-1.\nonumber
\end{eqnarray}
The last equation looks different for even and odd $N$:

\begin{eqnarray}
\label{eo}
\dot r_M &=& {2d-1\over 2d}Nr_1\left(2r_{M-1}-2r_M\right), \quad N=2M,\\
\dot r_M &=& {2d-1\over 2d}Nr_1\left(r_{M-1}-r_M\right), \quad N=2M+1.
\end{eqnarray}
We have two stationary solutions. The first is 

\begin{equation}
\label{react}
\bar r_1=\bar r_2=\ldots=\bar r_M=\bar r_0-{1\over (2d-1)N}
\end{equation}
which together with the normalization condition yields 

\begin{eqnarray}
\label{react2}
\bar r_0&=&{2d-2\over (2d-1)N^2}+{1\over (2d-1)N}, \\
\bar r_i&=&{(2d-2)\over (2d-1)N^2}, \quad {\rm for}
\quad i=1,\ldots, N-1. 
\end{eqnarray}
This solution describes the reactive steady state.  Note that $\bar
r_i\propto (d-1)$, implying a drastic difference between 1D and higher
dimensional systems.  In 1D, $\bar r_i=0$ corresponding to coarsening is
feasible, while for $d>1$ we have $\bar r_1>0$ implying to a reactive
steady state.  The second stationary solution
\begin{equation}
\bar r_1=0, \qquad \bar r_i\ne 0 \quad {\rm when}\quad i\ne 1
\end{equation}
corresponds to fixation; it is possible in arbitrary dimension. 

To figure out which of these two solutions actually appears in the long
time limit let us solve Eqs.~(\ref{main}).  To accomplish this we first
replace variables $t$ and $r_j(t)$ by

\begin{equation}
\tau={(2d-1)N\over 2d}\int_0^t dt'\, r_1(t')
\end{equation}
and

\begin{equation}
R_0(\tau)=r_0(t)-{1\over (2d-1)N},\quad
R_i(\tau)=r_i(t).
\end{equation}
In these variables, Eqs.~(\ref{main}) acquire a pure diffusion form

\begin{equation}
\label{diff}
{dR_j\over d\tau}=R_{j-1}-2R_j+R_{j+1}.
\end{equation}
In these equations index is defined modulo $N$ as previously.
Equivalently, we may treat $R_j(\tau)$ as a periodic function of $j$.
The initial condition reads

\begin{equation}
\label{initial}
R_{j}(0)=\cases{{1\over N^2}-{1\over (2d-1)N}, 
&$j\equiv 0({\rm mod}\,N)$;\cr
{1\over N^2}, &{\rm otherwise}.\cr}
\end{equation}
Solving (\ref{diff}) subject to (\ref{initial}) yields 

\begin{equation}
\label{sol}
R_i(\tau)={1\over N^2}-{1\over (2d-1)N}\sum_{j=-\infty}^{\infty}
e^{-2\tau}I_{i+Nj}(2\tau),
\end{equation}
where $I_n$ denotes the modified Bessel function of order $n$.  If the
variable $R_1(\tau)=r_1(t)$ remains positive, the modified time variable
$\tau$ behaves similarly to the original time variable $t$; in
particular, $\bar r_i=r_i(t=\infty)=R_i(\tau=\infty)$.  The latter
quantity is easily found (from the general properties of diffusion
equation) to be equal to the averaged initial value.  Thus
$R_i(\infty)={2d-2\over (2d-1)N^2}$, and therefore we recover the reactive
steady state of Eq.~(\ref{react2}).  On the other hand, if $R_1(\tau)$
becomes equal to zero at some moment $\tau_f$, this will be the end of
evolution as $\tau=\tau_f$ would imply $t=\infty$.  This case thus
corresponds to fixation: $\bar r_1=0$, $\bar r_i=R_i(\tau_f)>0$ for
other $i$.

Practically, it is convenient to determine the minimum of $R_1(\tau)$ in
the range $0<\tau<\infty$; if the minimum is negative, fixation does
happen.  It turns out that the minimum becomes negative for sufficiently
large $N$.  This allows us to keep only the dominant term from
the infinite sum (\ref{sol}), so

\begin{equation}
\label{solu}
R_1(\tau)={1\over N^2}-{e^{-2\tau}I_1(2\tau)\over (2d-1)N}.
\end{equation}
The minimum is reached at $\tau=\tau_*\cong 0.77256363$, and
$R_1(\tau_*)$ becomes negative when $N\geq 4.564293\times (2d-1)$.
Given $N$ should be integer it implies $N_c=14$ in 2D.  Would we keep
all terms in the sum, we would get a little smaller non-integer
threshold but still the same $N_c(2)=14$.  This assertion can be checked
numerically with great accuracy if we note that the sum in (\ref{sol})
can be significantly simplified.  Indeed, using the well-known
identity\cite{bender}

\begin{equation}
\label{identity}
\sum_{j=-\infty}^{\infty}z^j I_j(2\tau)=\exp[(z+z^{-1})\tau].
\end{equation}
one can derive
 
\begin{equation}
\label{id}
\sum_{j=-\infty}^{\infty}I_{1+Nj}(2\tau)={1\over N}
\sum_{p=0}^{N-1}\zeta^{-p}\exp[(\zeta^p+\zeta^{-p})\tau],
\end{equation}
with $\zeta=\exp(2\pi\sqrt{-1}/N)$.
Combining (\ref{sol}) and (\ref{id}) we arrive at

\begin{displaymath}
R_1(\tau)={1\over N^2}-{1\over (2d-1)N^2}
\sum_{p=0}^{N-1}{\exp[(\zeta^p+\zeta^{-p}-2)\tau]\over\zeta^{p}} 
\end{displaymath}
which involves only finite summation.  This expression has been used to
check that indeed $R_1(\tau)$ remains positive only for $N<14$ in 2D.
One can compute $N_c(d)$ in arbitrary dimension; for instance
$N_c=5,14,23,32,42,51$ when $d=1,2,3,4,5,6$, respectively.

\begin{figure}
\narrowtext
\epsfxsize=2.5in\epsfysize=2.5in
\hskip 0.3in\epsfbox{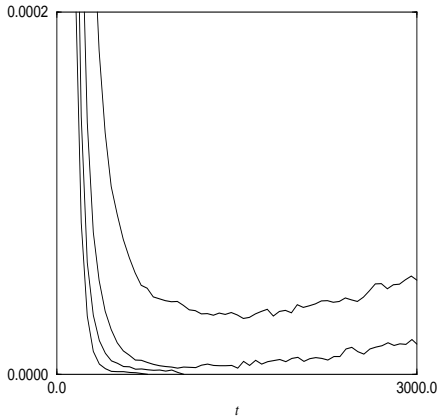}
\vskip 0.15in
\caption{Time dependence of the concentration of reactive pairs $r_1(t)$ 
in two dimensions.  Shown are Monte Carlo simulation results for
$N=12,13,14,15$ (top to bottom).
\label{fig1}}
\end{figure}

Thus we have found the critical number of opinions $N_c(d)$ within the
realm of Kirkwood approximation.  To determine actual $N_c$, numerical
simulations have been performed.  We have considered 2D square lattices
(maximum size $2048\times 2048$) with periodic boundary conditions.  On
lattice of these sizes, fixation has been found for $N\geq 14$.  The
concentration of reactive pairs $r_1(t)$ is drawn on Fig.~1 for
$N=12,13,14,15$ (the simulation data represent an average over 20
different realizations).  Fig.~2 plots the concentration of reactive
pairs provided by the Kirkwood approximation.

\begin{figure}
\narrowtext
\epsfxsize=2.5in\epsfysize=2.5in
\hskip 0.3in\epsfbox{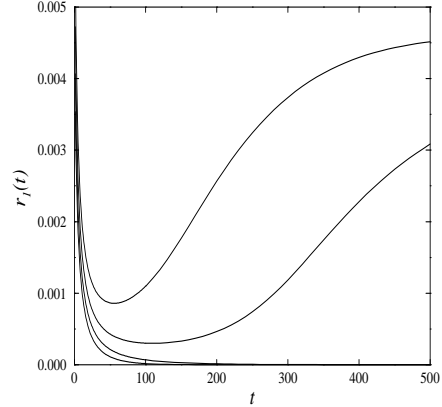}
\vskip 0.15in
\caption{Numerical integration of Eqs.~(9) in two dimensions.  
Shown are the concentrations of reactive pairs for $N=12,13,14,15$ (top
to bottom).
\label{fig2}}
\end{figure}

A word of caution is in order.  For large $N$, the concentration of
reactive pairs saturates at a very small value.  Statistical
fluctuations around this value may drive this value to zero, which is an
absorbing state.  This would imply an apparent fixation.  Even for
sufficiently large systems a few samples with $N=13$ have reached
this absorbing state.  However, the role of fluctuations reduces with
size, and for linear sizes of order $256$ and higher we have typically
seen a reactive steady state when $N=13$.  In contrast, fixation has
always been observed for $N=14$ for linear sizes up to $2048$.  Strictly
speaking, our numerical results provide a lower bound for the threshold
value: $N_c\geq 14$.  However, present data support much stronger
assertion $N_c=14$, identical to our theoretical prediction based on the
Kirkwood approximation.

To demonstrate the validity of the Kirkwood approximation it is
instructive to apply it to the cyclic voter model in 1D where a variety
of results were already established\cite{bramson,f,lpe}.  For $N=3$ and 
$d=1$ we solve rate equations to find
\begin{equation}
\label{N=3t}
r_1(t)=r_2(t)={1\over 9}\,{1\over 1+t/2}.
\end{equation}
Similarly, for $N=4$ and $d=1$ we find
\begin{eqnarray}
\label{N=4t}
r_1(t) &=& r_3(t)={1\over 16}\,{1\over 1+t/2},\nonumber\\
r_2(t) &=& {1\over 8}\,{1\over \sqrt{1+t/2}}
-{1\over 16}\,{1\over 1+t/2}.
\end{eqnarray}
Thus in both cases the Kirkwood approximation predicts $1/t$ decay of
the density of reactive interfaces.  The long time behaviors for $N=3$
and $N=4$ cyclic voter model in 1D agree with our previous mean-field
results for these cases\cite{lpe}.  Compare to exact results\cite{lpe},
however, mean-field treatments predict faster kinetics; e.g., the
density of reactive interfaces decays as $t^{-1/2}$ and $t^{-2/3}$ for
$N=3$ and 4, respectively\cite{lpe}.  As for the threshold number, both
rigorous approaches and mean-field treatments give the same value
$N_c(1)=5$.  This suggests that $N_c(d)$ given by the Kirkwood
approximation might be exact in higher dimensions as well.

In summary, we investigated the cyclic lattice Lotka-Volterra model.  We
argued that for sufficiently large number of species, $N\geq N_c$,
fixation occurs.  Within the framework of Kirkwood approximation, the
threshold value $N_c(d)$ has been found analytically in arbitrary
dimension; for instance, $N_c=5,14,23,32,42,51$ when $d=1,2,3,4,5,6$.
In one dimension this prediction is exact and in two dimensions it
agrees with our numerical findings for lattices of size up to $2048\times
2048$.

\vskip 0.18in
\noindent 
We thank E.~Ben-Naim and R.~Zeitak for helpful
discussions.  This research was supported in part by the Swiss National
Foundation, the ARO (grant DAAH04-96-1-0114), and the NSF (grant
DMR-9632059).

\end{multicols} 

\end{document}